\documentstyle[preprint,12pt,aps]{revtex}
\tightenlines
\begin{document}
%
\title{Polydisperse star polymer solutions}
\author{C. von Ferber, A. Jusufi, M. Watzlawek,\footnote{%
Present address: Bayer AG, Central Research Division,
D-51368 Leverkusen, Germany} C. N. Likos, and H. L\"owen}
\address{Institut f\"ur Theoretische Physik II,
Heinrich-Heine-Universit\"at 
D\"usseldorf,\\ 
Universit{\"a}tsstra{\ss}e 1, D-40225 D\"usseldorf, Germany
}
\maketitle
\begin{abstract}
  We analyze the effect of polydispersity in the arm number on the
  effective interactions, structural correlations and the phase
  behavior of star polymers in a good solvent. The effective
  interaction potential between two star polymers with different
  arm numbers is derived using scaling theory. The resulting
  expression is tested against monomer-resolved molecular dynamics
  simulations. We find that the theoretical pair potential is in
  agreement with the simulation data in a much wider polydispersity
  range than other proposed potentials. We then use this pair
  potential as an input in a many-body theory to investigate
  polydispersity effects on the structural correlations and the phase
  diagram of dense star polymer solutions. In particular we find that
  a polydispersity of 10\%, which is typical in experimental samples,
  does not significantly alter previous findings for the phase diagram
  of monodisperse solutions.
  \\
  PACS {82.70.Dd, 64.60.Fr, 61.20.Ja, 61.41.+e}
\end{abstract}
\date{\today}
\section{Introduction}\label{I}

A star polymer consists of linear polymer chains tethered to one
common central core. The number of polymer chains $f$ is usually
referred to as {\it arm number} or {\it functionality}.  If the degree of
polymerization, i.e., the number of monomers per chain, $N$, is the
same for all chains, the star polymer is called monodisperse
with respect to $N$ or  ``regular''. For large $N$, the size of
the central core particle is much smaller than the overall extension
of the whole star. For a single star, the density of monomers around
the central particle, $\phi(r)$, is radially symmetric, and from this
density profile the so-called corona diameter $\sigma$, which measures
the extension of the star, can be defined as the diameter of a sphere
around the star center where all $Nf$ monomers are found ``inside''
\cite{Daoud:Cotton:82:1}.

In the last years, star-shaped polymer aggregates have attracted a
considerable amount of interest from both the experimental and the
theoretical point of view
\cite{Grest:review:96,Burchard:review:99,Freire:review:99}.  The
reason for that is threefold. First, from a technical point of view,
star polymers are important for several industrial applications
\cite{Grest:review:96}.  One example are hydrogenated polyisoprene
star polymers, which are used as viscosity index modifiers in oil
industry applications due to their excellent shear stability.
Further, commercial star polymers are brought into action in coating
materials, as binders in toners for copying machines, and in several
pharmaceutical and medical applications \cite{Grest:review:96}.
Second, from an experimental point of view, the recent synthesis of
regular star polymers with various possible numbers of arms by Roovers
and coworkers \cite{Zhou:93:1,Roovers:93:1} made it possible to
explore the physics of well-defined model systems, which are
monodisperse in both the number of arms and the degree of
polymerization.  Important examples are polyisoprene stars with
$f=8,18$ \cite{Dozier:91:1} and polybutadiene stars with $f=32,64,128$
\cite{Roovers:93:1}, both synthesized by anionic polymerization.
Third, star polymers constitute an important soft-condensed matter
system, linking the fields of polymer physics and colloid physics,
thus attracting also interest from a purely theoretical point of view.
Star polymers with small arm numbers ($f=1,2$) resemble linear
polymers.  Thus, their configurations show a
considerable asphericity \cite{Solc:71:1,Solc:71:2,Tanaka:96:1},
although their chain-averaged number density of monomers, $\phi(r)$, is
spherically symmetric around the center of mass of the polymer.  With
increasing arm number $f$, the asphericity of the stars has been shown
to decrease considerably
\cite{Solc:73:1,Mattice:80:1,Batoulis:Kremer:89:1,Jagodzinski:94,Zifferer:95:1,Zifferer:97:1,Zifferer:99:1,Forni:97:1,Sikorski:98:1},
leading to ``stiff'' spherical particles in the limit of high $f$. It
is essentially this limit that a description of star polymers as
sterically stabilized colloidal particles holds.  This polymer-colloid
hybrid character of star polymers has been explored in a number of
publications dealing with the structural
\cite{Grest:review:96,Dozier:91:1,Witten:Pincus:86:1,Witten:Pincus:86:2,Richter:93:1,Roovers:97:1,Adam:91:1,Gast:review:96,Willner:92:1,Willner:94:1,Likos:stars:98:1,Watzlawek:98,Martin_PRL,martin:phd:00,Ferber:Jusufi:99:1}
and dynamical
\cite{Roovers:97:1,Adam:91:1,Gast:review:96,Roovers:94:1,Seghrouchni:98:1,Stellbrink:97:1,Vlassopoulos:97:1,Nommensen:98:1,Fleischer:preprint:1}
properties of star polymers.

In the usual theoretical description, star polymers in dense solution
are considered as an effective one-component (monodisperse) system,
i.e., all the stars have the same arm number $f$ and all the linear
polymer chains attached to the center have the same molecular weight.
While the latter can be realized by a careful preparation method, the
former is in general not true in actual samples.  The preparation of star
polymers proceeds by adding linear polymer chains to a dendrimer-like
core with reactions centers placed at the end of the dendrimer and the
polymer chains.  The chemical reaction is typically incomplete such
that few of the reaction centers at the dendrimer are not linked to a
linear chain. Hence an intrinsic polydispersity in the arm number
arises.  By ultracentrifugation
one can estimate the relative polydispersity in the arm number to be
around 5-15 percent in carefully and slowly prepared samples. It can,
however, also be much larger for fast reactions
\cite{Japanese,Japanese2}.  In a comparison between experimental data
and theory, in almost all previous studies the assumptions of a
monodisperse sample was implicitly made such that a natural question
concerns the influence of polydispersity on the statistical properties
of star polymer solutions. As far as theory is concerned, the
situation resembles much the case of colloidal suspensions where the
effect of charge and size polydispersity has been a topic of intense
recent research, 
see Refs.\ \cite{polydis,polydis2,polydis3,polydis4,polydis5,polydis6}
for recent reviews.

In this paper, we investigate the effect of arm number polydispersity
on the effective interaction, the structural correlations and the
phase diagram for star polymer solutions in a good solvent.  Our work
is based on a theoretical analysis using scaling theory and computer
simulations.  We first derive the effective interaction between two
star polymers of different arm numbers $f_1$ and $f_2$ from scaling
assumptions. Basically the interaction is logarithmic with the
core-core distance $r$ between the two stars but the prefactor depends
explicitly on $f_1$ and $f_2$ and differs from that of a monodisperse
sample. The resulting pair potential is tested against molecular
simulations and good agreement is found even when $f_1$ is very
different from $f_2$ while earlier descriptions \cite{Lomba} are found
to hold for small polydispersities but to be too simplistic for large
polydispersities.  In a further step towards a full description of
dense star polymer solutions, we then perform computer simulations of a
classical many-body systems interacting by means of this effective pair
interaction. We find that a polydispersity in the arm number reduces
the structural correlations but this effect is less drastic than for
size polydispersity in sterically stabilized colloidal suspensions.
Further, we present
evidence that the reentrant melting behavior predicted
from the theoretical treatment of a monodisperse description
\cite{Witten:Pincus:86:2,Martin_PRL,martin:phd:00,McConnell:97:1} 
does not change drastically in a
polydisperse solution of star polymers.

The paper is organized as follows: In section \ref{II} we present the
scaling ideas and derive the effective interaction potential between
two stars of different arm numbers.
A test of this
result against molecular computer simulation is performed in section
\ref{III}. Results for the structure and phase diagram from a simulation of
a many-body system are given in section \ref{IV}. Finally, we conclude in
section \ref{V}.

\section{Scaling theory for an effective potential}\label{II}
The effective interaction between two star polymers results from the 
steric interaction of all monomers that constitute the polymer chains
on each of the stars. It is obviously a formidable task to derive
the effective potential from first principles. Luckily we are in the 
position to give a rigorous result for the limiting case of short 
distance between the two star cores, while good arguments exist for the 
description of the long range part of the interaction. We have found 
in the treatment of the interaction of monodisperse star polymers that
the combination of these two approaches leads to good agreement with 
simulation results as well as with scattering experiments on real 
star polymer systems \cite{Likos:stars:98:1,me,Likos2}.

We first derive the short distance interaction using scaling
arguments.  Many details of the behavior of polymer solutions may be
derived using the renormalization group (RG)
analysis \cite{Schaefer99}. Here, we use only the
more basic results of power law scaling: the radius of gyration
$R_1(N)$ of a single linear polymer chain with $N$ monomers and its
partition function ${\cal Z}_1(N)$ are found to obey the power laws:
\begin{equation}\label{1}
R_1(N)\sim N^\nu \mbox{\hspace{3em} and \hspace{3em}}
{\cal Z}_1(N)\sim z^N N^{-\nu\eta_2} \,.
\end{equation}
The fugacity $z$ measures the mean number of possibilities to add one
monomer to the chain. It is microscopic in nature and will depend on
the details of the theoretical model or 
the experimental system.  The two exponents
$\nu$ and $\eta_2$ on the contrary are universal to all polymer
systems in a good solvent, i.e., excluding high concentration of
polymers or systems in which the polymers are collapsed or are near
the collapse transition.  A family of additional exponents $\eta_f$
governs the scaling of the partition functions ${\cal Z}_f(N)$ of 
polymer stars of $f$ chains each with $N$ monomers while the power law
for the radius of gyration remains unchanged:
\begin{equation}\label{2}
R_f(N)\sim N^\nu \mbox{\hspace{3em} and \hspace{3em}}
{\cal Z}_f(N)\sim z^{fN} N^{\nu(\eta_f-f\eta_2)} .
\end{equation}
The exponents of any other power law for more general polymer networks
are given by scaling relations in terms of $\eta_f$ and $\nu$
\cite{Duplantier86,Ohno88,Schaefer92}. Note that 
eqs.\ (\ref{1}) and (\ref{2}) imply
$\eta_1=0$.  The values of these exponents are known from
RG and Monte Carlo (MC) simulations
\cite{Batoulis89}.  In the limiting case of many arms, $f\gg 1$, the leading
behavior of $\eta_f$ is given by \cite{Ohno89}
\begin{equation}\label{3}
\eta_f\sim -f^{3/2},
\end{equation}
a result that has been found also from of geometrical considerations using
the blob model of a star polymer \cite{Witten:Pincus:86:1}.
The scaling law for the partition sum of two star
polymers may be derived from a short distance expansion
\cite{Duplantier89,Ferber97}. The partition sum of the two stars
${\cal Z}_{f_1f_2}(N,r)$ at small center-to-center
distances $r$ factorizes into a function $C_{f_1f_2}(r)$
and the partition function ${\cal Z}_{f_1+f_2}(N)$ of the
star with $f_1+f_2$ arms that is formed when the cores of the two
stars coincide:
\begin{equation}\label{4}
{\cal Z}_{f_1f_2}(N,r)\sim C_{f_1f_2}(r)
{\cal Z}_{f_1+f_2}(N)\,.
\end{equation}
Such a behavior is generally assumed in scaling arguments .  At large
star-star separation ${\cal Z}_{f_1f_2}(N,r)$ is the product of the single
star partition functions. Again we may argue that the ratio of the
partition function of two stars at finite separation $r$ to that at
infinite separation can only be a function $B$ of $r/R_1$ noting that
$R_1$ is the only relevant length scale in this problem ($R_{f_1}$ and
$R_{f_2}$ differ from $R_1$ only in an $f$-dependent prefactor):
\begin{equation}\label{5}
{\cal Z}_{f_1f_2}(N,r)\sim B_{f_1f_2}(r/R_1)
{\cal Z}_{f_1}{\cal Z}_{f_2}\,.
\end{equation}
Taking this into account and inserting the power law scaling according
to Eqs. (\ref{1}) and (\ref{2}) we find that Eqs. (\ref{4}) and
(\ref{5}) are only compatible if also $C(r)$ follows a power law
\begin{eqnarray}
C_{f_1f_2}(r)&\sim &r^{\Theta^{\rm (s)}_{f_1f_2}}\,,
\end{eqnarray}
and the so-called contact exponent $\Theta^{\rm (s)}_{f_1f_2}$ obeys the 
scaling relation
\begin{eqnarray}\label{6}
\Theta^{\rm (s)}_{f_1f_2} &=&  \eta_{f_1}+\eta_{f_2}-\eta_{f_1+f_2}\,.
\end{eqnarray}
Note that in terms of the RG of polymer field theory the exponents
$\eta_f$ correspond to dimensions of operators associated with the
partition functions of single stars \cite{Schaefer92} and relation
(\ref{6}) is a direct consequence of the short distance expansion.
The mean force $F_{f_1f_2}(r)$ acting on the centers of two star
polymers with $f_1$ and $f_2$ arms is now derived as the gradient of
the effective potential $V^{\rm eff}(r)= -k_{\rm B}T\log[ {\cal
  Z}_{f_1f_2}(r)/({\cal Z}_{f_1}{\cal Z}_{f_2})]$ with $k_{\rm B}T$
denoting the thermal energy. For the force at short distances $r$ this
evidently results in
\begin{equation}\label{7}
 F_{f_1f_2}(r)=k_{\rm B}T 
  \frac{\Theta^{\rm (s)}_{f_1f_2}}{r}\,.
\end{equation}
Using the above many arm limit for $\eta_f$ one may match the contact
exponents $\Theta^{\rm (s)}_{ff}$ to the known values for $f=1,2$
\cite{desCloizeaux75} fixing the otherwise unknown prefactor in eq.
(\ref{3}). Assuming that the behavior of the $\Theta^{\rm (s)}_{ff}$
may be described by this approximation for all $f$ one finds:
\begin{equation}
 F_{ff}(r)\approx k_{\rm B}T\frac{5}{18} \frac{f^{3/2}}{r} \,.
\end{equation}
This matching in turn suggests an approximate value for the $\eta_f$
exponents, taking into account $\eta_1=0$ and $\eta_2\approx -5/18$:
\begin{equation}\label{8}
 \eta_f\approx -\frac{5}{36} \frac{f^{3/2}-f}{\sqrt{2}-1} \,.
\end{equation}
Inserting this into the general formula for the interaction of two
different star polymers the contact exponent reads for large $f_1$ and
$f_2$
\begin{equation}\label{9}
 \Theta^{\rm (s)}_{f_1f_2}= 
 \frac{5}{36}\frac{1}{\sqrt{2}-1}
 \left[\left(f_1+f_2\right)^{3/2}-\left(f_1^{3/2}+f_2^{3/2}\right)\right].
\label{theta}
\end{equation}

We note that on a phenomenological level there are other possibilities
to describe the interaction in a polydisperse system of star polymers.
In colloidal systems two other approaches to the interaction of
polydisperse particles are commonly used: in a steric stabilized
system with a polydispersity in the radii of the particles one expects
to find an interaction radius that is the {\em arithmetic mean} of the
radii of the two interacting particles \cite{polydis4,barrat}.  In a
charge stabilized solution the other hand, the interaction of two
particles is proportional to the product of the two charges and the
effective charge is calculated as the {\em geometric mean} of the two
charges.  The latter description was used in an earlier investigation
of star polymers treating $f$ as an ``effective charge'' parameter
\cite{Lomba,daguanno:klein:92,elshad:98}.  For comparison with these
other possible approaches we define:
\begin{equation}\label{10}
 \Theta^{\rm (a)}_{f_1f_2}(r)= 
 \frac{5}{18}\left(\frac{f_1+f_2}{2}\right)^{3/2}
\,,
\end{equation}
\begin{equation}\label{11}
 \Theta^{\rm (g)}_{f_1f_2}(r)= 
 \frac{5}{18}\left(\sqrt{f_1f_2}\right)^{3/2}
\,.
\end{equation}

We now turn to the interaction at larger separation.  It has been
recently shown that a Yukawa like tail for the potential between star
polymers reproduces the results of both simulation and scattering
experiments for monodisperse solutions
\cite{Likos:stars:98:1,me,Likos2}.  The natural scale for a star
polymer is the corona diameter $\sigma_f=2\lambda R_f$ with
$\lambda\approx 2/3$ \cite{me}.  For the potential between two $f$-arm star
polymers at distance $r>\sigma_f$ one has
\begin{equation}
V_{ff} (r) \sim \frac{1}{r}\exp(-r\kappa_f).
\end{equation}
The decay length $1/\kappa_f=2\sigma_f/\sqrt{f}$ is the diameter of
the outermost blobs of the star polymer in the Daoud-Cotton model
\cite{Daoud:Cotton:82:1,Likos:stars:98:1}.

In a simple but successful approach the full potential for the
interaction between two $f$-arm star polymers was constructed by
concatenating the long range and short range potentials.  In this
model the crossover from short to long range behavior takes place when
the distance of the two cores is $\sigma_f$.  It is natural to
generalize this to two star polymers with $f_1$ and $f_2$ arms in such
a way that the crossover point is at distance
$\sigma=(\sigma_{f_1}+\sigma_{f_2})/2$.  It is less obvious how to
choose the decay length of the long range part.  Our choice
$1/\kappa=1/\kappa_{f_1}+1/\kappa_{f_2}$ is in accordance to the
simulation results presented below. Note that although we have omitted
the subscripts on $\sigma$ and $\kappa$, the latter are functions of
$f_1$ and $f_2$.

The full potential then reads
\begin{equation}\label{13}
\frac{1}{k_{\rm B}T}V_{f_1f_2}(r)=\Theta^{\rm (s)}_{f_1f_2}\left\{
\begin{array}{ll}
\displaystyle -\ln(\frac{r}{\sigma})
 +\frac{1}{1+\sigma\kappa}
 &\displaystyle {\rm for}\;\; r\leq \sigma;\\
\displaystyle  \frac{1}{1+\sigma\kappa} \frac{\sigma}{r}
\exp(\sigma\kappa-r\kappa) &\displaystyle {\rm for}\;\; r > \sigma.\\
\end{array}
\right.
\end{equation}
In the monodisperse case $f_1=f_2$ this potential reduces to the one
between two identical star polymers that has been successfully tested
in extensive simulations.

\section{Comparison with simulation results}\label{III}
We have performed Molecular Dynamics (MD) computer simulations of two
star polymers with different arm numbers $f_{1}$ and $f_{2}$ by
resolving the monomers as classical beads along the chains.  We use
the same simulation model as in previous studies of
star polymer solutions in a good solvent \cite{me,Grest}.
The arm length, i.e., the
number $N$ of beads along each chain, was kept fixed to $N = 50, 100$
which was shown in previous work to be large enough to guarantee a
sufficient scaling behavior.  The main features of the simulation
model can be summarized as follows:
\begin{enumerate}
\item A purely repulsive  truncated Lennard-Jones interaction
between all $N(f_{1}+f_{2})$ monomers is assumed.  
\item An attractive FENE-potential is added for the interaction  between
  neighboring mo\-no\-mers along a chain.
\item To accommodate the polymer arms, a hard core with radius $R^{(d)}$ is
introduced at the
  center of the stars. The cores interact with the monomers according
  to the above mentioned potentials with a separation shift of
  $R^{(d)}$.      
\end{enumerate}
Two stars with arm numbers $f_1$ and $f_2$ were fixed at core
separation $r$.  After a long equilibration period, the total force
$F_{f_1f_2}$ acting onto the cores is averaged. The temperature $T$ is kept
constant.  Both the equilibration time and the time during which
averages were performed was about $1000\tau$ where $\tau
=\sqrt{m\sigma^2 / \epsilon}$ is the Lennard-Jones time unit with
$\sigma$, $\epsilon$ and $m$ denoting the Lennard-Jones length, energy
scale and the monomer mass.  A simulation snapshot is shown 
in Fig.\ \ref{fig0}.
When comparing the mean force to the derivative of the
theoretical effective potential, two technical problems arise: (i) In
contrast to the theory, we have a finite core size
$R^{(d)}$ which is the same for both stars.  (ii) The two corona
diameters, $\sigma_{f}$, ($f=f_1,f_2$) which enter in the potential of
eq.\ (\ref{13}), are not directly accessible in a simulation.

As in the monodisperse case \cite{me}, the first difficulty (i) is
overcome by plotting the inverse force $1/F_{f_1f_2}$ versus distance $r$. The
divergence in the force occurs already at core separations
$r \cong 2R^{(d)}$ which leads to a zero in the $1/F_{f_1f_2}$ plot. The
inverse {\it slope} of the $1/F_{f_1f_2}$-plot then corresponds to the
prefactor $\Theta^{(s)}_{f_1f_2}$ of the logarithmic term in the
effective pair potential. It is this slope which can directly be
compared to our theoretical prediction eq. (\ref{13}).  Furthermore, a
linear function in the $1/F_{f_1f_2}$ plot is a direct check for the
validity of the $\ln r$-term in the effective interaction potential,
eq.\ (\ref{13}).

To handle the second difficulty (ii) we consider the radius of
gyration $R_{f}$ of each star which is readily accessible in a
simulation of a single star\cite{me}. We assume a proportionality
between the $R_{f}$ and $\sigma_{f}$,
\begin{equation}
\sigma_{f} = 2 \lambda R_{f}
\end{equation}
and use $\lambda$ as a fit parameter to fit the full force versus
distance curve for given arm numbers $f_1$ and $f_2$. The results for
the optimal fit parameter are shown in Table \ref{parameters.table}.
We obtain an averaged value of $\langle\lambda\rangle \approx 0.66$,
which is nearly independent of $f_{1}, f_{2}$. Thus, the results found
here are consistent with those from the previously investigated
monodisperse case \cite{me}. We further remark that this value
coincides with that used in 
Ref.\ \cite{Likos:stars:98:1} to fit experimental data
for $f=18$
and that $\lambda$ is
independent of $N$, consistent with scaling theory.

Results for the comparison between theory and simulation are shown in
Figure \ref{invforce.plot} where the reduced inverse mean force
$1/F_{f_1f_2}$ between the star centers is plotted versus intercore
distance $r$ for two arm numbers $f_{1} = 10$ and $f_{2} = 50$. We
observe two important facts: first, the data indeed fall on a straight
line proving the logarithmic behavior of the potential inside the mean
corona. The straight line hits the origin at $r=2R^{(d)}$ showing the
relevance of the finite core in the simulations.  Second, the inverse
slope $\Theta_{f_1f_2}$ agrees very well with the theoretical
prediction also shown in Figure \ref{invforce.plot}. A similar
behavior was observed for all other combinations of arm numbers
contained in Table \ref{parameters.table}.
 
In Fig. \ref{prefactor.plot}, the prefactor $\Theta_{f_1f_2}$ is
plotted versus $f_{2}$, at a fixed value $f_{1}=10$, and compared to
the different theoretical predictions in eqs. (\ref{9})-(\ref{11}).  Of
course, in the monodisperse case $f_{1}=f_{2}=10$, all the different
theoretical expressions for $\Theta_{f_1f_2}$ coincide and agree well
with the simulation, consistently with our earlier work \cite{me}. For
increasing asymmetry between $f_{1}$ and $f_{2}$, significant
deviations between the predictions of (a) an arithmetic or (g) a geometric
mean and the simulation results become visible.  While the
expression $\Theta_{f_1f_2}^{{\rm (s)}}$ gained from scaling theory is able
to describe the slope even for large asymmetries,
$\Theta_{f_1f_2}^{{\rm (g)}}$ is worse at large asymmetries while
$\Theta_{f_1f_2}^{{\rm (a)}}$ even possesses the wrong curvature as a
function of $f_2$. Similar conclusions hold for other combinations of
$f_{1}$ and $f_{2}$.

In Fig. \ref{results.plot}, we show the force $F_{f_1f_2}$ versus
distance $r-R^{(d)}$, scaled by $R_{12}=(R_{f_1}+R_{f_2})/2$,
comparing the theoretical force with simulation results for three
cases: $f_{1} = 10, 18, 30; f_{2} = 50; N = 50$.  Note that formally
the core size $R^{(d)}$ vanishes in the theory.  
There is a good overall agreement inside
the corona region between theory and simulation even for large
asymmetries of $f_{1}$ and $f_{2}$.  Results for distances $r$ outside
the corona, $r>\sigma$, are presented in Figure \ref{lnforce.plot}.
Here, we test the exponential decay of the force $F_{f_1f_2}$ with
distance $r$ by plotting the logarithm of the force versus distance.
The crossover of the inner-core data to a straight line outside the
core is clearly visible. Within the simulation error bars, the slope
is consistent with the theoretical prediction eq. (\ref{13}).


\section{Polydispersity effects in dense star polymer solutions}\label{IV}
 
To explore the effects of arm number polydispersity on the structure
and phase behavior of dense solutions of star polymers, we further
performed large scale Monte Carlo simulations using the pair potential
approach as given by the potential in eq.\ (\ref{13}).  Thus, on the
one hand, this approach closely corresponds to our recent work on the
structural and phase behavior of monodisperse star polymer solutions
\cite{Watzlawek:98,Martin_PRL,martin:phd:00}, and, on the other side,
is quite similar to investigations of polydispersity effects in
solutions of hard spheres \cite{polydis4,barrat,kofke1,kofke2} and
charged colloidal particles
\cite{polydis5,daguanno:klein:92,elshad:98,klein:daguanno} available
in the literature.

In our MC simulations, representative particle configurations of
approximately 2000 star polymers in a cubic simulation box were
created in the following manner.  First, a starting particle
configuration was built up by placing the particles on random
positions in the box, each particle assigned with an arm number $f$
chosen from a Gaussian distribution $g(f)$ of mean $\bar f$ and
variance $p = \sqrt{{\overline{f^2}} - {\bar f}^2}$:
\begin{equation}
g(f) = \frac{2}{p\sqrt{2\pi}}
\exp\left[-\frac{1}{2 p^2}\left(f - \bar f\right)^2\right].
\label{gaussian}
\end{equation}
We use eq.\ (\ref{gaussian}) above to {\it define} the polydispersity
$p$ in what follows.  The Gaussian distribution chosen to describe arm
number polydispersity agrees well with experimentally-determined arm
number distributions of star polymers synthesized by anionic
polymerization \cite{Likos2}.  Note that nonphysical, negative arm
numbers, which are in principle not prohibited by a Gaussian arm
number distribution, did not occur in the simulations for the values
of $\bar{f}$ and $p$ chosen here.  Second, an equilibration phase of
approximately $50\,000$ MC cycles was performed by allowing both
translational particle moves according to the standard Metropolis
scheme, and particle exchanges of randomly chosen particle pairs,
again using the Metropolis rule to decide an exchange to happen or
not.  For the calculation of pair potential energies, eq. (\ref{13})
was used, together with the scaling relation \cite{Daoud:Cotton:82:1}:
\begin{equation}
   \label{scaling.diameters}
   \frac{\sigma_{f_1}}{\sigma_{f_2}}=\left(\frac{f_1}{f_2}\right)^{1/5}.
\end{equation}
After the equilibration phase, approximately $100\,000$ MC cycles were
simulated to gather statistics for both the radial center-to-center
distribution function $g(r)$ and the center-to-center structure factor
$S(k)$ \cite{polydis,polydis2,polydis4,polydis5,hansen:mcdonald}.  In
what follows, we will use $\bar{\sigma}=\sigma_{\bar{f}}$ as the basic
length scale, and the mean packing fraction $\bar{\eta}= \rho
\bar{\sigma}^3\pi/6$ ($\rho$ being the number density of the stars) as
a measure of the density.

In order to investigate the polydispersity effects 
on the liquid structure of dense star polymer solutions,
simulations with varying polydispersities 
$p$ were performed at (average) arm numbers
$\bar{f}$ and mean packing fractions $\bar{\eta}$, 
known to correspond to the liquid state
in the monodisperse case $p=0$
\cite{Martin_PRL,martin:phd:00}.
We show typical results for $g(r)$ and $S(k)$ in fig.\ \ref{gs.f32}.
As can be seen, increasing the polydispersity leads to 
decreasing spatial correlations in the
fluid, indicated by a decreasing principal 
peak height of both $g(r)$ and $S(k)$.
Thus, as expected, the effect of arm number 
polydispersity in star polymer solutions is
quite similar to the effect of charge polydispersity 
in solutions of charged colloids
\cite{polydis5,klein:daguanno}.
Note that the anomalous behavior of 
$S(k)$ reported for monodisperse star polymer solutions
in Ref.\ \cite{Watzlawek:98} 
also holds for polydisperse star polymer solution 
for all polydispersities $p\le 14$ simulated.

In order to explore the evolution of the freezing and reentrant
melting phase transitions found for monodisperse star polymer
solutions \cite{Martin_PRL,martin:phd:00} with increasing
polydispersity, we employed the following strategy.  We performed a
number of simulations at state points corresponding to the solid phase
in the monodisperse case, gradually increasing the parameter $p$.  In
particular, we chose $\bar{f}=40$ and $\bar{\eta}=0.5$ in a first set
of simulations.  Our results for $g(r)$ and $S(k)$ are given in Fig.\ 
\ref{gs.f40}.  Let us begin with the discussion of the $g(r)$-data.
As the monodisperse case $p=0$ corresponds to an solid state of
bcc-symmetry \cite{Martin_PRL,martin:phd:00}, the shown $g(r)$ is, in
fact, the radially-averaged pair correlation function of the stars in
the crystal state.\footnote{In fact, from the typical
  particle-particle distances read off from the pair correlation
  function, it can be concluded that the crystal structure is bcc, a
  result which is in agreement with calculations of bond order
  correlation functions \cite{martin:phd:00}.}  Again, increasing $p$
leads to decreasing structural correlations between the particles
indicated by a decreasing principal peak height and a ``smearing-out''
of the sub peak-structure of $g(r)$ seen for $p=0$. This scenario also
manifests itself in the structure factors shown in fig.\ 
\ref{gs.f40}(b).  Notice that the corresponding structure factor for
$p=0$ exhibits strong Bragg peaks, again indicating the monodisperse
system to be in the crystal state for $\bar{f}=f=40$,
$\bar{\eta}=\eta=0.5$, and is not depicted for that reason.  For
nonzero polydispersities the simulated structure factors do not show
Bragg peaks and the value of the principal peak heights $S_{\rm max}$
decreases.  Thus, increasing the polydispersity in a sample of star
polymers at $\bar{f}=f=40$, $\bar{\eta}=\eta=0.5$ leads to a melting
transition from a $bcc$ crystal to a fluid phase.  We therefore expect
polydisperse star polymer solutions to show an enlarged fluid phase
region as compared to the phase diagram of monodisperse star polymers
in Refs.\ \cite{Martin_PRL,martin:phd:00}.  Furthermore, the stability
range of the fluid phase is expected to increase with increasing the
polydispersity $p$.  To corroborate this expectation in more detail,
we have performed further MC simulations of polydisperse star polymers
for various state points ($\bar{f}$, $\bar{\eta}$), all corresponding
to state points close to the freezing or reentrant melting line in the
monodisperse case \cite{Martin_PRL,martin:phd:00}.  For the ranges of
polydispersity examined here, ($p \leq 14$), no change in the topology
of the phase diagram and no new crystalline phases were found as a
result of polydispersity.  A quantitative calculation of the full
phase diagram by more sophisticated simulation methods
\cite{kofke1,kofke2} is beyond the scope of this paper and will be
left for future studies.

 
\section{Conclusions}\label{V}

In conclusion, we have analyzed the effect of arm number
polydispersity on the effective interaction and the structural
correlations between star polymers in a good solvent. An analytical
expression for an effective pair potential, given by eq.\ (\ref{13}),
was put forward and its validity was confirmed by Molecular Dynamics
computer simulation. This pair potential was subsequently used to
simulate the structural correlations between many stars.  As expected,
correlations decrease with increasing polydispersity.  At the same
time, though, the effect is much less pronounced than for
size-polydisperse hard spheres. This can be seen in analogy to
colloidal suspensions where size polydispersity of
sterically-stabilized suspensions is known to lead to much more
pronounced effects than charge-polydispersity in charged suspension.
The microscopic reason for that is that the effective interaction is
much softer for charged colloids which is similar to our case of star
polymers.

We finish with a couple of remarks: first it would be interesting to
compare, both qualitatively and quantitatively, our theoretical
predictions to experimental data. In fact, the intrinsic
polydispersity in the arm number can be measured and structural
correlations are accessible, e.g., by neutron scattering of a star
polymer solution with marked cores. Second, it would be interesting to
map out, for a given polydispersity, the full phase diagram including
freezing into different solid structures.  Also, it would be
interesting to develop and apply a liquid integral equation theory to
predict structural correlations in a polydisperse star polymer
solutions. The output of such a theory could be checked against our
computer simulation data of section \ref{IV}. Third, we have not considered
a polydispersity in the length of the linear polymer chains attached
to the centers.  It would be interesting to study this theoretically
and compare the results to samples prepared in such a way that
different linear chains are brought to the reaction centers of the
dendrimers. Work along these lines is left for future studies.

\acknowledgments

This work has been supported in part by the SFB 237 of the Deutsche
Forschungsgemeinschaft.


\begin{table}[hbt]
\begin{center}
   \begin{tabular}{|c|c|c|c|}\hline

     $f_{1}$ & $f_{2}$ & $N$ & $\lambda$\\\hline

     3 & 5 & 100 & 0.511 \\

     5 & 10 & 100 & 0.643\\

     5 & 18 & 100 & 0.667\\

     10 & 18 & 50 & 0.675\\

     10 & 30 & 50 & 0.673\\

     10 & 50 & 50 & 0.692\\

     18 & 30 & 50 & 0.696\\

     18 & 50 & 50 & 0.668\\

     30 & 50 & 50 & 0.714\\ \hline

   \end{tabular}

 \end{center}
 \caption{ \label{parameters.table} List of the simulated
     $f_{1}$-$f_{2}$-combinations and the corresponding results for
     $\lambda=\sigma/(R_{f_1}+R_{f_2})$. 
     $N$ is the associated monomer number per arm.}
\end{table}   
%
\begin{figure}[hbt]            
  \caption{ \label{fig0} 
Snapshot of two interacting star polymers with $f_1=18$ and $f_2=10$ arms.}
\end{figure} 
\begin{figure}[hbt]
  \caption{ \label{invforce.plot} 
    Reduced inverse force $k_{B}T/(R_{12}F_{f_1f_2})$
    between the centers of two star polymers (for $f_{1}=10, f_{2}=50$
    and $N=50$) versus reduced distance $r/R_{12}$,
    where $R_{12} = (R_{f_1} + R_{f_2})/2$. The error bars
    were obtained by averaging over the results of $10$ independent
    simulations. The dashed line is a linear regression of the data.
    The solid line is the theory from chapter \protect\ref{II}.}
\end{figure} 
\begin{figure}[hbt]
   \caption{ \label{prefactor.plot} Simulation results for the
            prefactor $\Theta_{f_1f_2}$ for $f_{2}=5\ldots 50$
            and $f_{1}=10$ fixed, in comparison to three different
            theoretical predictions.}
\end{figure}
\begin{figure}[hbt]            
  \caption{ \label{results.plot} 
            Simulation results (symbols) and theoretical results (lines) 
            for the reduced effective force $R_{12}F_{f_1f_2}/k_{B}T$ versus
            reduced distance $(r-2R^{(d)})/R_{12}$.}
\end{figure}
\begin{figure}[hbt]            
  \caption{\label{lnforce.plot}Logarithm of the reduced force,
    $\ln\left(R_{12}F_{f_1f_2}/k_{B}T\right)$, versus reduced
    distance, $(r-2R^{(d)})/R_{12}$, for $f_{1}=10, f_{2}=18$ and
    $N=50$.  The error bars were obtained by averaging over the
    results of $10$ independent simulations.  }
\end{figure} 
\begin{figure}[hbt]
\caption{
   \label{gs.f32}
   (a) Radial distribution function $g(r)$ and 
   (b) center-to-center structure factor $S(k)$ for
   polydisperse star polymers in the fluid state. 
   In the MC simulations, the average arm number
   has been chosen as $\bar{f}=32$, 
   the mean packing fraction as $\bar{\eta}=0.5$.
   Polydispersities $p$ as indicated in the figure.
}
\end{figure}


\begin{figure}[hbt]
\caption{
   \label{gs.f40}
   (a) Radial distribution function 
   $g(r)$ and (b) center-to-center structure factor $S(k)$ for
   $\bar{f}=40$, $\bar{\eta}=0.5$.
   Polydispersities $p$ as indicated in the figure.
}
\end{figure}

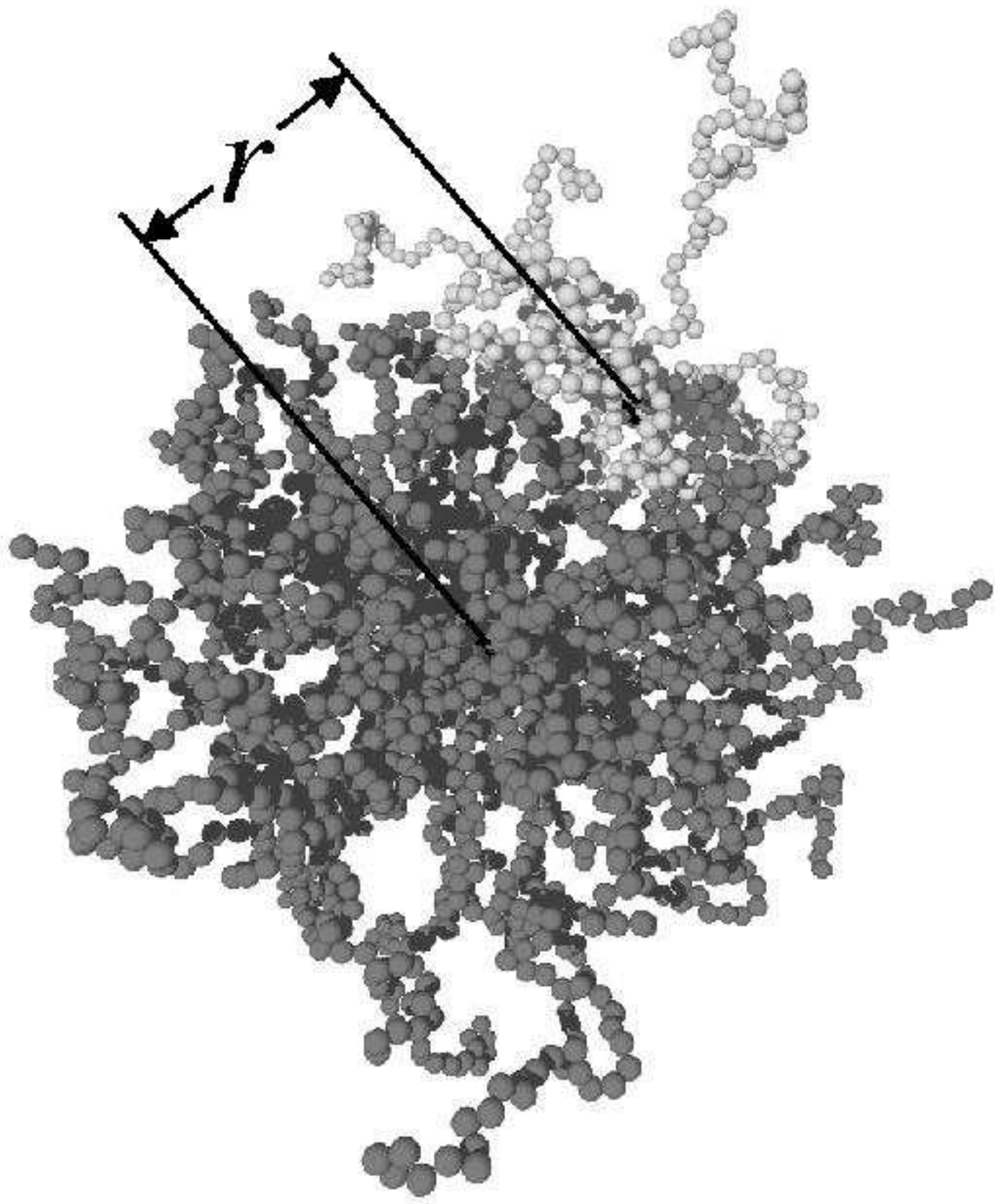
\end{document}